\documentclass[reprint, superscriptaddress, amsmath, amssymb, aps]{revtex4-1}
\usepackage{epstopdf}
\usepackage{graphicx}
\usepackage{dcolumn}
\usepackage{bm}
\usepackage{multirow}
\usepackage[table,xcdraw]{xcolor}
\usepackage{color,soul}
\begin{document}
\title{First-principles study on the solute-induced low diffusion and self-trapping of helium in fcc iron}
\author{Kui Rao}
\author{Jingxin Hu}
\affiliation{Department of Physics and Key Laboratory for Low-Dimensional Structures and Quantum Manipulation (Ministry of Education), Hunan Normal University, Changsha 410081, China}
\author{Gang Ouyang}
\author{Zi-Ran Liu}
\email{zrliu@hunnu.edu.cn}
\affiliation{Department of Physics and Key Laboratory for Low-Dimensional Structures and Quantum Manipulation (Ministry of Education), Hunan Normal University, Changsha 410081, China}
\author{Xinfu He}
\email{hexinfu@ciae.ac.cn}
\author{Wen Yang}
\affiliation{Division of Reactor Engineering Technology Research, China Institute of Atomic Energy, Beijing 102413, China}

\begin{abstract}
The addition of alloying elements plays an essential role in helium (He) behaviours produced by transmutation in metal alloys.
Effects of solutes (Ni, Cr, Ti, P, Si, C) on the behaviours of He and He-He pair in face-centred cube (fcc) iron have been investigated using first-principles calculations based on density functional theory (DFT). For the interactions of solutes and He, we found that Ti, P, Si, and C attracts He is more potent than Ni and Cr in fcc iron. We have determined the most stable configuration for the He-He pair, which is the He$_{sub}$-He$_{tetra}$ pair with a binding energy of 1.60 eV. In considering the effect of solutes on the stability of the He-He pair, we have proposed a unique definition of binding energy. By applying the definition, we suggest that Ti and P could weaken He self-trapping, and Cr and C are beneficial for He self-trapping, while Ni is similar to the matrix Fe itself. For the diffusion of He, which is the necessary process of forming the He bubble, we determined that the most stable interstitial He is in a tetrahedral site and could migrate with the energy barrier of 0.16 eV in pure fcc iron. We further found that Ti and Si can increase the barrier to 0.18 and 0.20 eV; on the contrary, Cr and P decrease the barrier to 0.10 and 0.06 eV, respectively. Summarizing the calculations, we conclude that Ti decreases while Cr increases the diffusion and self-trapping of He in fcc iron.
\end{abstract}

\maketitle

\section{Introduction}
He can be produced in the structural materials by nuclear transmutation ($\emph{n}$, $\alpha$) through neutron irradiation. Due to its low solubility, He can be deeply trapped in lattice defects, such as dislocations \cite{heinisch2006interaction,Wang2008EffectsOE,zhu2015helium}, grain boundaries \cite{Zhang2013EnergeticLA,Kurtz2004TheEO,ishiyama1996post}, precipitates \cite{cao2018first,He2014FirstprinciplesSO}, especially vacancy and vacancy clusters \cite{li2013vacancy,zhang2014he,Zhang2011StabilityAM,Hepburn2013FirstPS,fu2005ab,yang2008ab,You2017BubbleGF}, inducing bubble formation and void swelling, which would cause high-temperature embrittlement of materials \cite{trinkaus2003helium,Veen2003HeliumII,Kesternich2003MechanicalPA,zhang2015properties}.

Austenitic steels, being widely used in current fission reactors as structural alloys in the internal components, which will be so for the foreseeable future because of their relatively high strength, ductility, and fracture toughness \cite{chopra2011review,Zinkle2013MaterialsCI}. However, void swelling and He embrittlement is unavoidable due to the microstructure of material evolving in a severe irradiation environment. He plays an important role in this microstructural evolution \cite{ullmaier1984influence,Maziasz1984SwellingAS}. Many researchers have devoted their efforts to overcoming the problems of swelling and embrittlement in austenitic steels. One of the practicable proposals is introducing precipitates within the matrix as trapping centers for He, which has been proven feasible to improve swelling and He embrittlement in materials \cite{Maziasz1984SwellingAS,Kesternich1985APS,shiraishi1979effects,kesternich1981reduction,du2017effects,
Lee1984TheEO,lee1986mechanism,garner1985role,boothby1988effects,Maziasz1984SwellingAS,kimoto1985void,
maziasz1993void,lee1992relationships}.

Solutes Ti, C, P, and Si will form various precipitates in austenitic steels during irradiation, which can effectively disperse and capture He, reducing the size of He bubbles in grain boundaries. Previous experimental works showed that TiC precipitates in austenitic steels is efficient for trapping He due to their finely dispersed distributions in the matrix, so that the precipitates can reduce the He accumulates at grain boundaries, and TiC precipitates is also supposed to pin the dislocation. As a result, the addition of titanium is beneficial in reducing the sensitivity of irradiation temperature to high-temperature ductility of materials \cite{Maziasz1984SwellingAS,Kesternich1985APS,shiraishi1979effects,kesternich1981reduction,du2017effects}. It was found that the interface of phosphide particles and matrix serves as a site for the nucleation of a fine dispersion of He bubbles, and phosphorus increases the diffusivity of matrix solvent atoms, reducing the vacancy supersaturation during irradiation \cite{Lee1984TheEO,lee1986mechanism,garner1985role}. Similarly, the addition of silicon in austenitic steels can form (Ni, Si)-rich precipitates that can also capture He, and the silicon enhances diffusivity of the alloy, which leads to a substantial increase in the free energy barrier to void nucleation \cite{boothby1988effects,garner1981effect}.

About the magnetic states in fcc iron, experiments demonstrated that the magnetic structure of the $\gamma$-Fe is a spin-density wave (SDW) state in which the noncollinear spiral spin configuration is a possible structure \cite{tsunoda1989spin}, which has been simulated using local spin-density functional approximation and linear-muffin-tin-orbitals approach \cite{uhl1992electronic,mryasov1992spiral,mryasov1991magnetic,Krling1996GradientcorrectedAI}. Thus, the spiral spin-density wave state is an ideal magnetic structure of $\gamma$-Fe, but it is a challenge to investigate $\gamma$-Fe with such spiral spin magnetic configuration in current density functional theory (DFT) packages. However, it seems feasible to solve the modelling problem by applying localized magnetic moments in paramagnetism state of $\gamma$-Fe. Some first-principles studies have shown that it is possible to use a set of ordered collinear magnetic structures to model the $\gamma$-Fe, and many magnetic states have been explored \cite{Krling1996GradientcorrectedAI,herper1999ab,herper1999ab,jiang2003carbon,Klaver2012DefectAS,
Hepburn2013FirstPS,piochaud2014first}. Especially, Hepburn et al. applied localized moments structures to model paramagnetism state of Fe-Cr-Ni austenitic steels successfully by first-principles \cite{Klaver2012DefectAS,Hepburn2013FirstPS,piochaud2014first}. It is predicted that the antiferromagnetic double layer (afmD) collinear magnetic structure has the lowest energy per Fe atom among all the collinear magnetic states \cite{jiang2003carbon,Klaver2012DefectAS}, which is consistent with experimental results and the calculations using the full potential linear-augmented plane-wave method \cite{herper1999ab}.

The behaviour and interactions of He in bcc and fcc metals have been studied by using DFT methods \cite{Zhang2017EffectOC,Zhang2011StabilityAM,Hepburn2013FirstPS,fu2005ab,yang2008ab,You2017BubbleGF,zeng2009first,
Li2016EffectsOC,Zhang2015HeHeAH,Liao2020FirstprinciplesSO,Zhang2017FirstprinciplesSO,
Li2019FormationAM,Becquart2006MigrationEO,Cao2016MigrationOH,Zu2009PropertiesOH,
You2018TheBO,Deng2013DiffusionOS,Niu2014WaltzingOA}. As an initial stage for He clustering, self-trapping plays an important role in the formation of He bubbles in metals. The binding energy is the main physical quantity for describing self-trapping. The binding energies of tetrahedral (tetra)-tetrahedral (tetra) He pairs were 0.43 eV in bcc Fe \cite{fu2005ab}, 0.60 eV in Al \cite{yang2008ab}, 0.97 eV in Mo \cite{Zhang2015HeHeAH}, 1.03 eV in W \cite{Becquart2006MigrationEO}, about 0.10 eV in V, Nb and Ta, and 0.70 eV in Cr \cite{Zhang2015HeHeAH}, respectively. While the energies for substitutional (sub)-interstitial (int) He pairs are more than 1.60 eV in both bcc Fe and Ni \cite{fu2005ab, Liao2020FirstprinciplesSO}. The interaction between solutes and He is also an object of attention. In bcc Fe, previous work suggested that many elements exhibit attraction to He, and the solute-He$_\emph{sub}$ pairs show much stronger attraction than that of solute-He$_\emph{tetra}$ pairs \cite{You2017BubbleGF}. However, the interactions between solutes and He are rarely investigated in fcc Fe.

\begin{figure}[h]
\center\includegraphics[scale=0.32]{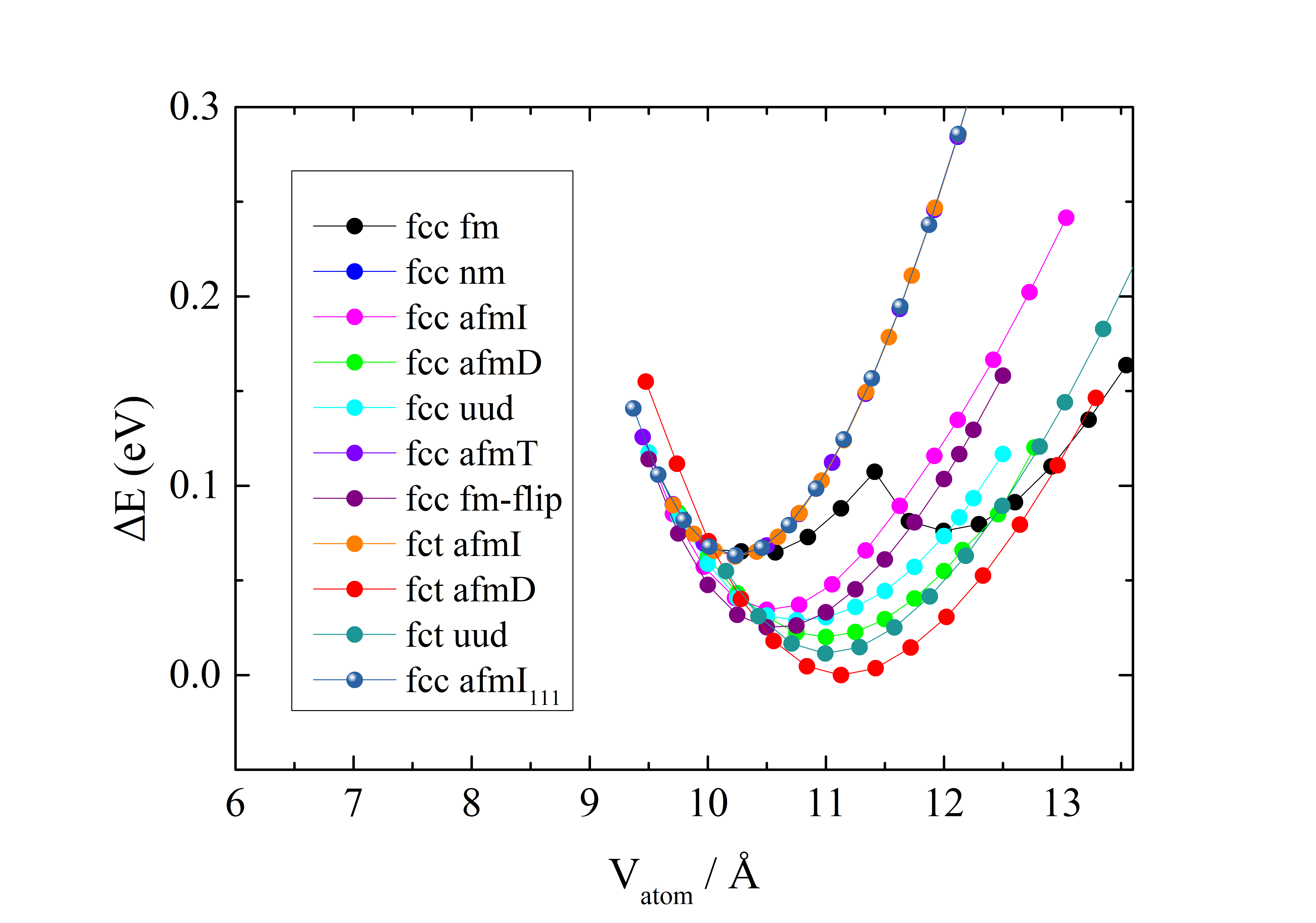}
\caption{The energy difference per atom of various collinear magnetic structures of fcc Fe relative to the most stable collinear magnetic state one, where fct means that the lattice parameters of unit cell is a=b$\neq$c. The mentioned magnetic states are shown in supplementary materials Fig. S1.} \label{fig0}
\end{figure}

He diffusion is related to the growth of the He bubble in metal materials. Previous ab initio calculations have shown that the migration energy barriers of a single interstitial He atom were 0.04 eV in Pt \cite{Cao2016MigrationOH}, 0.06 eV in bcc Fe \cite{Li2016EffectsOC}, 0.07 eV in W \cite{Becquart2006MigrationEO} and V \cite{Zhang2011StabilityAM}, 0.08 eV in Cu \cite{Cao2016MigrationOH}, 0.10 eV in fcc Al \cite{yang2008ab}, 0.13 eV in Ni \cite{Hepburn2013FirstPS}, and 0.15 eV in Pd \cite{Cao2016MigrationOH}.
Besides, small He clusters can migrate at low temperatures in bcc Fe with the migration energy barrier of 0.09 $\pm$ 0.02 eV for interstitial He-He pair, which was studied by molecular dynamics (MD) \cite{Deng2013DiffusionOS}, and the migration of interstitial He-He pairs in other bcc and fcc metals were also investigated by using first-principles density function calculations \cite{Cao2016MigrationOH,Niu2014WaltzingOA}.
Therefore, it is necessary to know the migration energy barrier of He in fcc iron, and also that of He with alloying elements.

An atomic-scale understanding of the effect of primary and minor alloying elements (Ni, Cr, Ti, P, Si, C) on behaviours of He is important for designing new austenitic steels since He plays a vital role in the evolution of microstructures. In the present work, firstly, we have calculated the interactions between two He atoms in which we determined the most stable He-He pair in fcc iron, and the interactions between solute and He in fcc iron were also studied. Based on the most stable He-He pair, solute-He-He complexes were investigated to obtain the binding energy of solute with He-He pairs, as well as the effects of solute on He self-trapping in fcc iron. Finally, solute effects on interstitial He migration energy barriers in fcc Fe have been calculated, and the conclusion was drawn. The calculations would be helpful to reveal the He diffusion and self-trapping at an early age in austenitic steels under irradiation.


\section{Computational details}
All related calculations were performed based on DFT as implemented in the Vienna Ab initio Simulation Package (VASP) with projector-augmented wave (PAW) potential \cite{Kresse1996EfficientIS,Blchl1994ProjectorAM,Kohn1965SelfConsistentEI}. The electron exchange-correlation functional was described within the generalized gradient approximation using PW91 functionals \cite{Perdew1992AtomsMS} and spin interpolation of the correlation potential provided by the improved Vosko-Wilk-Nusair scheme \cite{Vosko1980AccurateSE}. The tests (Fig. S2) show that the cut-off energy was set as 400 eV is precise enough. A $2\times2\times2$ \emph{k}-point Monkhorst-Pack grid was used to sample the Brillouin zone for supercells with 256 atoms. For the relaxation of single configuration, the ionic force on each atom is set as 10$^{-2}$ eV/{\AA}. In order to compute the energy barrier, the climbing image nudged elastic band (CINEB) method was used to find the transition pathways, which is a small modification to the NEB method in which the highest energy image is driven up to the saddle point \cite{Henkelman1999ADM}. In the CINEB calculation, the ionic force on each atom is set as 10$^{-1}$ eV/{\AA}. All calculations employed spin-polarization to account for the ferromagnetic state of fcc Fe and solute Ni.

The formation energy of a substitutional (sub) He atom, $E_{f}$, which can be defined as:
\begin{eqnarray}\label{eq1}
E_{f}({\rm He}_{sub})&=&E({\rm Fe}_{255}{\rm He})-\frac{255}{256}E({\rm Fe}_{256})\\ \notag
&&-E({\rm He}),
\end{eqnarray}
and the formation energy of an interstitial (int) He atom as:
\begin{eqnarray}\label{eq2}
E_{f}({\rm He}_{int})=E({\rm Fe}_{256}{\rm He})-E({\rm Fe}_{256})-E({\rm He}),
\end{eqnarray}
in which, $E$(Fe$_{256}$) is the energy of perfect supercell with 256 Fe atoms, $E({\rm Fe}_{255}{\rm He})$ is the energy of substituting one Fe atom by He in the supercell, $E({\rm Fe}_{256}{\rm He})$ is the energy of the supercell with the insertion of one interstitial He atom, $E({\rm He})$ is the energy of an isolated He atom in vacuum.

We define the binding energy between $\emph{n}$ atoms as:
\begin{eqnarray}\label{eq3}
E_{b}(A_{1},A_{2},\cdot\cdot\cdot,A_{n})&=&E(A_{1})+E(A_{2})+\cdot\cdot\cdot\\\notag
&&+E(A_{n})-E(A_{1},A_{2},\cdot\cdot\cdot,A_{n})\\\notag
&&-(n-1)E({\rm Fe}_{256})\notag,
\end{eqnarray}
where $E(A_{1}),E(A_{2}),\cdot\cdot\cdot,E(A_{n})$ are the energy of the supercell containing a single solute atom, $E(A_{1},A_{2},\cdot\cdot\cdot,A_{n})$ is the energy of the supercell containing $\emph{n}$ solute atoms. And the binding energy between a solute atom and He-He pair, can be calculated as:
\begin{eqnarray}\label{eq4}
E_{b}(A_{n},{\rm He}-{\rm He})&=&E(A_{n})+E({\rm He}_{1},{\rm He}_{2})\\ \notag
&&-E(A_{n},{\rm He}_{1},{\rm He}_{2})-E({\rm Fe}_{256}),
\end{eqnarray}
where, ${\rm He}_{1}$ and ${\rm He}_{2}$ refer to a single He atom in the supercell of different positions, $E({\rm He}_{1}$, ${\rm He}_{2}$) is the energy of the supercell containing a He-He pair, $E(A_{n}, {\rm He}_{1}, {\rm He}_{2}$) refers to the energy of the supercell containing $A_{n}-{\rm He}_{1}-{\rm He}_{2}$ complex. Specially, the binding energy of He-He pair when a solute exists in the supercell can be calculated as:
\begin{eqnarray}\label{eq5}
E_{b}^{A_{n}}({\rm He}-{\rm He})&=&E(A_{n},{\rm He}_{1})+E(A_{n},{\rm He}_{2})\\\notag
&&-E(A_{n},{\rm He}_{1},{\rm He}_{2})-E(A_{n}),
\end{eqnarray}
where, $E(A_{n}, {\rm He}_{1})$ is the energy of the supercell containing $A_{n}-{\rm He}_{1}$ complex, $E(A_{n}, {\rm He}_{2})$ refers to the energy of the supercell containing $A_{n}-{\rm He}_{2}$ complex. By definition, positive binding energy indicates attractive interaction for all conditions.

\begin{figure}[h]
\center\includegraphics[scale=0.3]{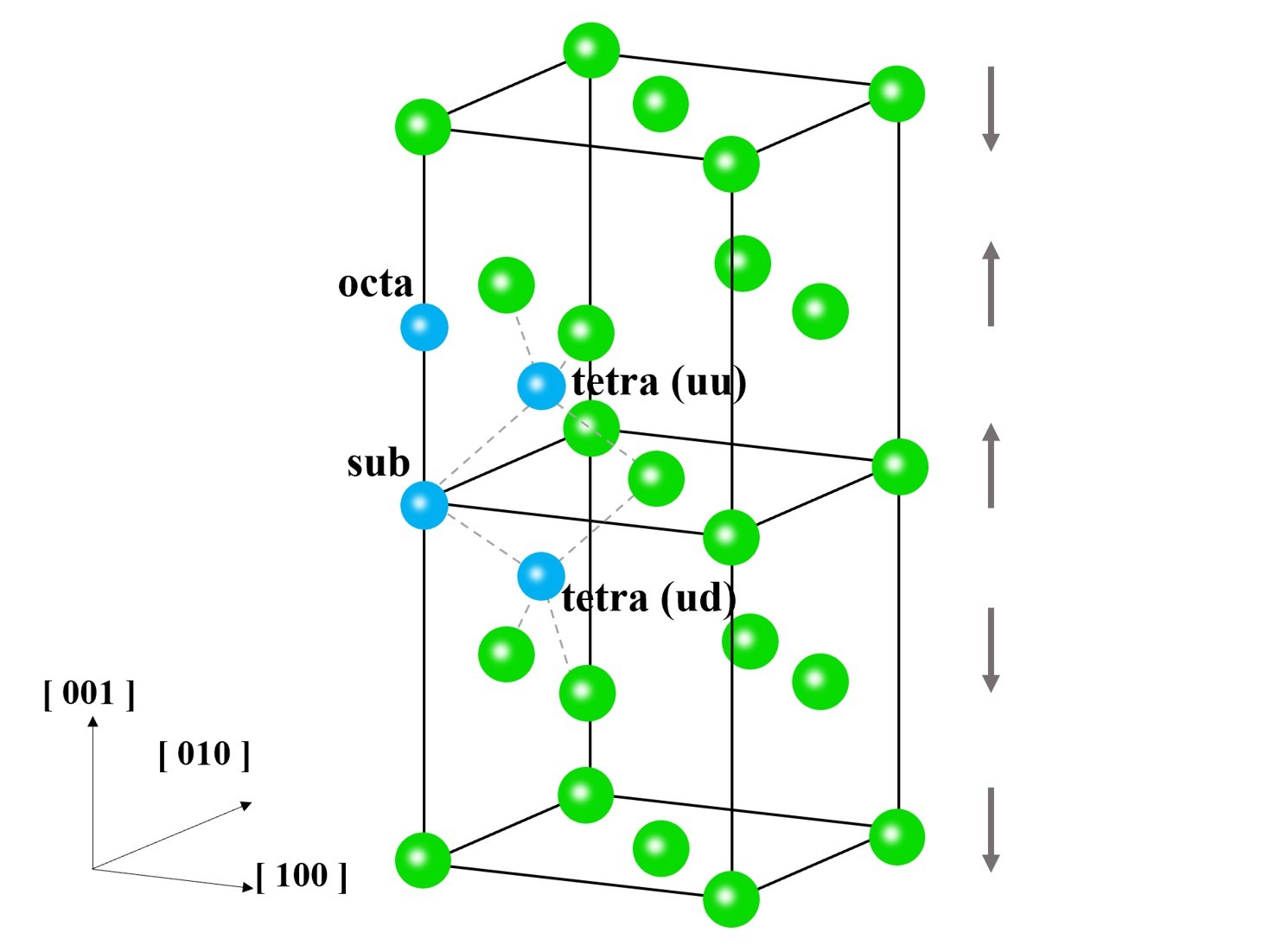}
\caption{Illustration of defects in fct afmD Fe. Green spheres refer to Fe atoms. Blue orbs refer to He atoms at substitutional (sub) site, interstitial octahedral (octa) and tetrahedral (tetra) site (uu refers to He in the site that located in up-up spin magnetic double-layers and ud refer to He in the site that located in up-down spin magnetic double-layers). Arrows indicate the local magnetic moments of Fe atoms one (001) layer.} \label{fig1}
\end{figure}

In order to obtain a suitable magnetic state for simulating fcc Fe, we apply many possible states, as shown in Fig. S1, and find that afmD collinear magnetic structure was the most stable state, which can be seen in Fig. \ref{fig0}. In the following calculations, we use fct afmD Fe as a matrix for all the simulations.

\section{He-He interactions}

A single He can occupy a substitutional or interstitial site, and the relatively stable one is the substitutional site, which is consistent with the  previous calculations in fcc Fe and studies of other bcc and fcc metals \cite{fu2005ab,Hepburn2013FirstPS, yang2008ab,zeng2009first,Zhang2017FirstprinciplesSO,Li2019FormationAM,
Becquart2006MigrationEO,Cao2016MigrationOH,Zu2009PropertiesOH,Liao2020FirstprinciplesSO,Zhang2017EffectOC}.
For interstitial defects, the tetrahedral site was more stable than the octahedral site. In detail, tetrahedral site ud is the most stable interstitial site because of the breaking symmetry by anti-ferromagnetic double-layer and its formation energy is 0.05 eV lower than tetrahedral uu and is 0.19 eV lower than octahedral site, respectively, configurations are shown in Fig. \ref{fig1}. As a closed-shell noble-gas element, He prefers sizeable free volume. That is why in fcc Al \cite{yang2008ab}, Pd \cite{zeng2009first,Zu2009PropertiesOH}, Pt \cite{Cao2016MigrationOH}, Ag \cite{Zu2009PropertiesOH} alloys, the octahedral site is more stable than the tetrahedral one. However, He$_\emph{tetra}$ is more stable than He$_\emph{octa}$ in fcc Fe due to the purely repulsive interactions of Fe-He \cite{Hepburn2013FirstPS}, though the volume of tetra-site is smaller than octa-site. A similar situation can be found in Cu \cite{Zu2009PropertiesOH} and Ni \cite{Hepburn2013FirstPS,Liao2020FirstprinciplesSO,Zu2009PropertiesOH}, which indicate that the tetrahedral site will be the most stable interstitial site in widely used Fe-Ni-based austenitic alloys.

\begin{figure}[h]
\center\includegraphics[scale=0.3]{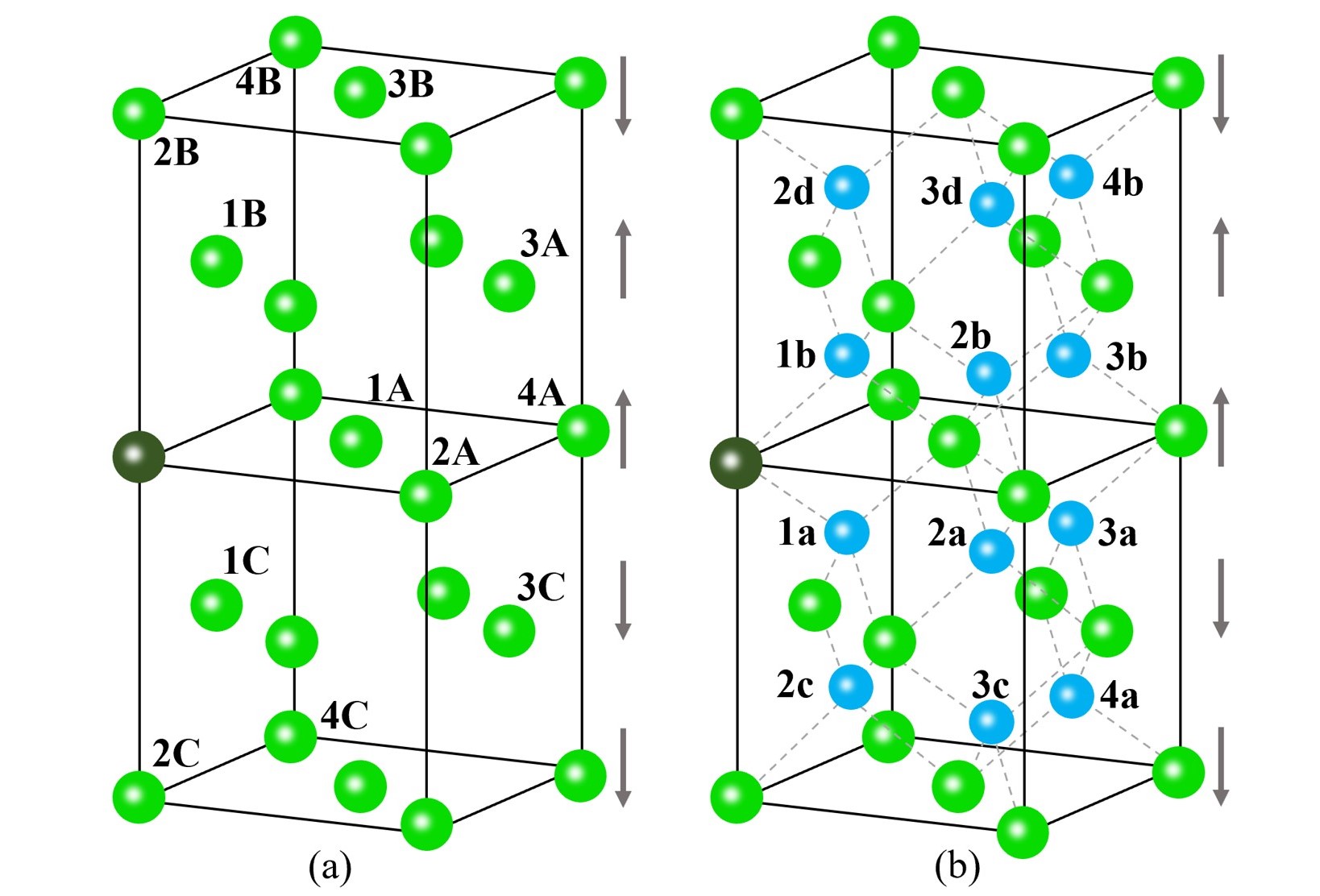}
\caption{Configurations for interactions between two foreign atoms in afmD Fe. (a) Dark green spheres refer to substitutional sites. (b) Dark green spheres refer to substitutional sites; blue spheres refer to tetrahedral cites. For the combination of numbers and letters in black font, numbers represent different nearest neighbour positions relative to dark green spheres and letters represent different positions of corresponding nearest neighbour.}   \label{fig2}
\end{figure}

As the initial stage for He clustering, self-trapping plays an important role in the formation and growth of He bubbles in Fe-based alloys \cite{Li2016EffectsOC}. We have performed calculations to investigate the interactions between pairs of He atoms for He$_\emph{sub}$-He$_\emph{sub}$, He$_\emph{tetra}$-He$_\emph{tetra}$ and He$_\emph{sub}$-He$_\emph{tetra}$ in fcc Fe, respectively, configurations are shown in Fig. \ref{fig2}.

\begin{figure}[h]
\center\includegraphics[scale=0.3]{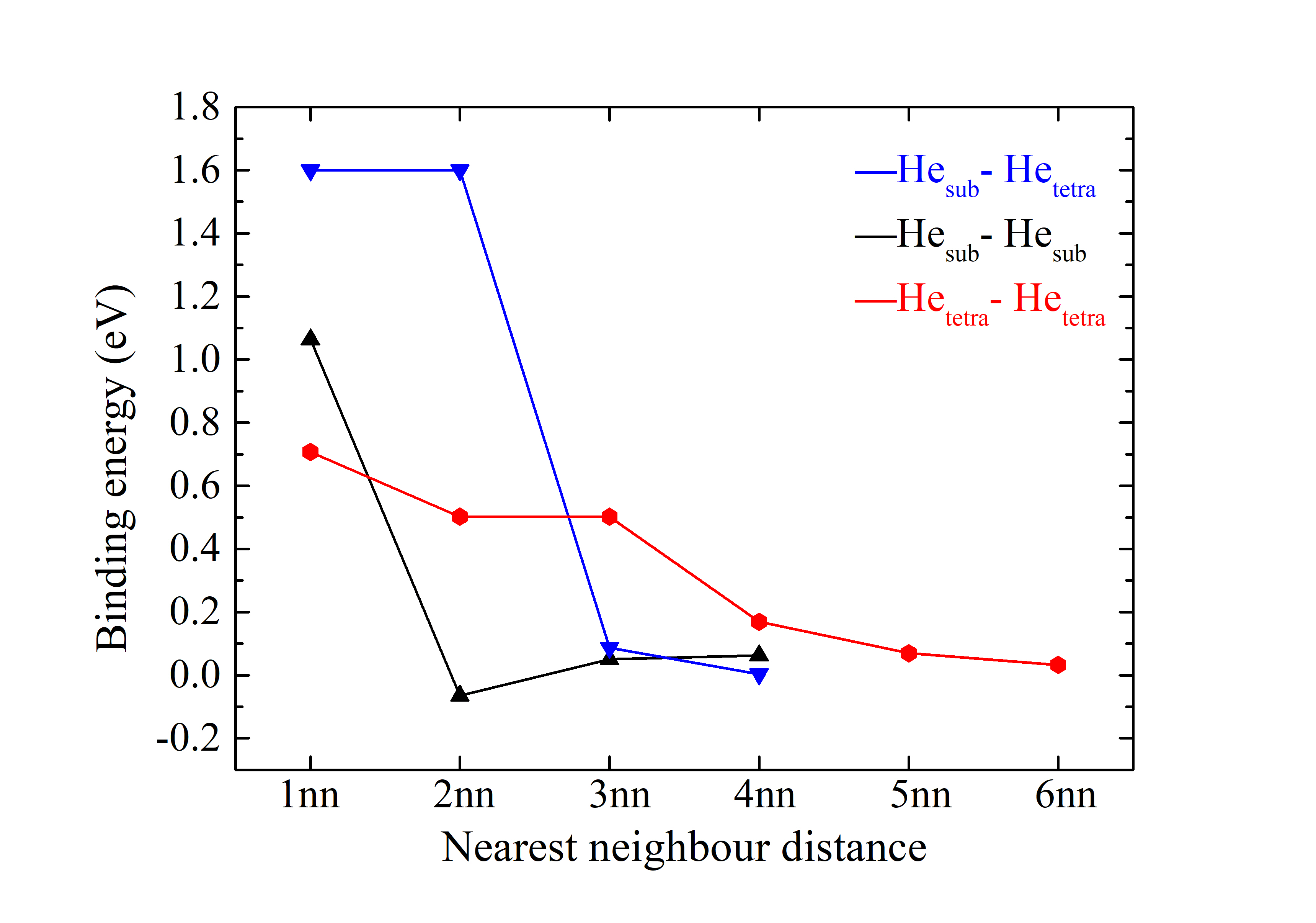}
\caption{The binding energy of He$_\emph{sub}$-He$_\emph{sub}$ pairs and He$_\emph{sub}$-He$_\emph{tetra}$ pairs of the separation distances from the first nearest neighbour (1 nn) to the fourth nearest neighbour (4 nn), and the binding energy of He$_\emph{tetra}$-He$_\emph{tetra}$ pairs of 1 nn to 6 nn.} \label{fig3}
\end{figure}

He$_\emph{sub}$-He$_\emph{sub}$ pairs exhibit relatively high binding energies at 1 nn with binding energy up to 1.06 eV and slightly repulsive interactions at 2 nn as shown in Fig. \ref{fig3}, and the interactions become very weak at 3 nn and 4 nn. In our calculations, He atoms at 1 nn relax directly towards each other by 0.65 to 0.90 {\AA}, resulting in He-He pairs with separation between 1.67 and 1.69 {\AA}, which meets the previous work \cite{Hepburn2013FirstPS}.

He$_\emph{tetra}$-He$_\emph{tetra}$ pairs show consistently attract in fcc Fe with positive binding energies at 1 nn, and He atoms are displaced slightly away from the tetrahedral sites under relaxation, resulting in He-He pairs with separation distances range from 1.62 to 1.68 {\AA}. For configurations of interactions between tetrahedral site and tetrahedral site which were shown in Fig. \ref{fig2}(b), three pairs of atoms which locate in the positions of 1 nn were considered. They are configurations of 2d-3d, 2d-1b and 1b-2b, three 2 nn configurations, that are 2d-4b, 2d-2b and 1b-3b, and the 3 nn configuration is 2d-3b, 4 nn configuration is 2d-1a, 5 nn configuration is 2d-2a and 6 nn configuration is 2d-3a, respectively.
The most stable He$_\emph{tetra}$-He$_\emph{tetra}$ pair is a configuration of 2d-3d with the binding energy of 0.71 eV, which is lower than the most stable He$_\emph{tetra}$-He$_\emph{tetra}$ pairs in Mo of 0.97 eV \cite{Zhang2015HeHeAH}, 1.03 eV in W \cite{Becquart2006MigrationEO}, higher than 0.43 eV in bcc Fe \cite{fu2005ab}, 0.60 eV in Al \cite{yang2008ab}, around 0.10 eV in V, Nb and Ta, and 0.70 eV in Cr \cite{Zhang2015HeHeAH}. For 2 nn He$_\emph{tetra}$-He$_\emph{tetra}$ pairs, the separation distances are reduced by 0.79 to 0.89 {\AA}, and considerable displacement for 3 nn is up to 1.43 {\AA}, resulting  in the distance between He-He atoms ranges from 1.63 to 1.65 {\AA} for 2 nn and 3 nn under relaxation. There is almost no displacement under relaxation for 4 nn to 6 nn He$_\emph{tetra}$-He$_\emph{tetra}$ pairs. However, still have a positive binding energy of about 0.03 eV at 6 nn,
indicating interstitial He may attract another interstitial He atom at least 4 {\AA} away, which benefits the formation of clusters.
This result also tells us that the following calculations on interactions between solutes and interstitial He atoms are important, and the diffusion of interstitial He in fcc Fe.

For He$_{sub}$-He$_{tetra}$ pairs, we found that all He$_{sub}$-He$_{tetra-1a}$, He$_{sub}$-He$_{tetra-1b}$, He$_{sub}$-He$_{tetra-2a}$, He$_{sub}$-He$_{tetra-2b}$ and He$_{sub}$-He$_{tetra-2c}$ directly relax to a substitutional vacancy containing two He atoms with the axis of He-He directions along $\langle1 1 \bar{1}\rangle$, $\langle111\rangle$, $\langle1 1 \bar{1}\rangle$, $\langle111\rangle$ and $\langle1 1 \bar{1}\rangle$, respectively.
There is a low or even no energy barrier for the transformation of 2 nn He$_{sub}$-He$_{tetra}$ pair to 1 nn pair. These pairs have the same He-He separation of 1.53 {\AA} with binding energy around 1.60 eV. Such a strong attraction between substitutional and tetrahedral He was also found in Ni with the binding energy of 1.51 eV for He$_{sub}$-He$_{tetra}$ pair \cite{Liao2020FirstprinciplesSO} and 1.84 eV in bcc Fe \cite{fu2005ab}. Moreover, we found that a substitutional vacancy with two tetrahedral He atoms, such as substitutional vacancy (dark green sphere in Fig. \ref{fig2}b) with He$_{tetra-1a}$ and He$_{tetra-1b}$, will relax to the same cluster as that of He$_{sub}$-He$_{tetra}$ pairs. We demonstrate that these He$_{sub}$-He$_{tetra}$ pairs will relax to a stable VHe$_{2}$ clusters (V stands for a vacancy) in fcc Fe with a binding energy of 3.93 eV. The He$_{sub}$-He$_{tetra}$ pairs of 3 nn and 4 nn, like He$_\emph{sub}$-He$_\emph{sub}$ pairs, exhibits weak binding energy with a maximum value of 0.09 eV.

\section{Solute-He interactions}
\begin{figure}[h]
\center\includegraphics[scale=0.3]{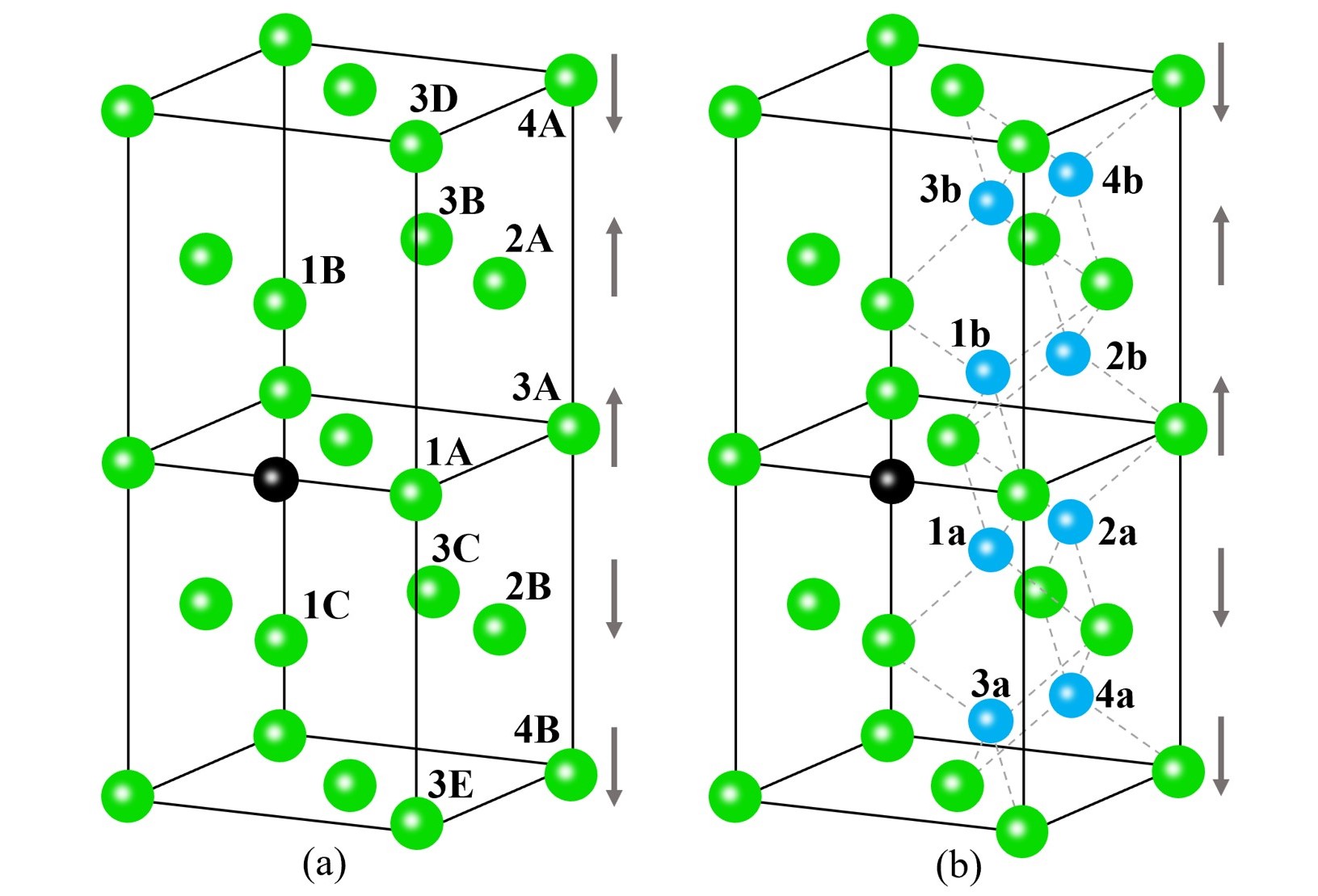}
\caption{Configurations for interactions between C atom (dark sphere) and substitutional/tetrahedral site He in afmD Fe. Green spheres refer to Fe atoms; dark spheres refer to octahedral C atom and blue spheres refer to interstitial sites of He. The numbers stand for the separation distances of substitutional site in (a) or interstitial site (blue sphere in (b)) relative to octahedral C atom of 1 nn to 4 nn.}  \label{fig4}
\end{figure}

According to our calculation, Ni, Cr, Ti, P, and Si prefer to occupy substitutional sites, while C atoms prefer octahedral sites with a formation energy of -8.60 eV. To figure out the interactions between these solutes and He, for solutes Ni, Cr, Ti, P, and Si, we use the configurations of solute-He as shown in Fig. \ref{fig2}, while for C, we apply the configuration of C-He as demonstrated in Fig. \ref{fig4}, in which 1 nn to 4 nn nearest neighbor were calculated.

\begin{figure*}[tbp]
\center\includegraphics[width=0.95 \textwidth]{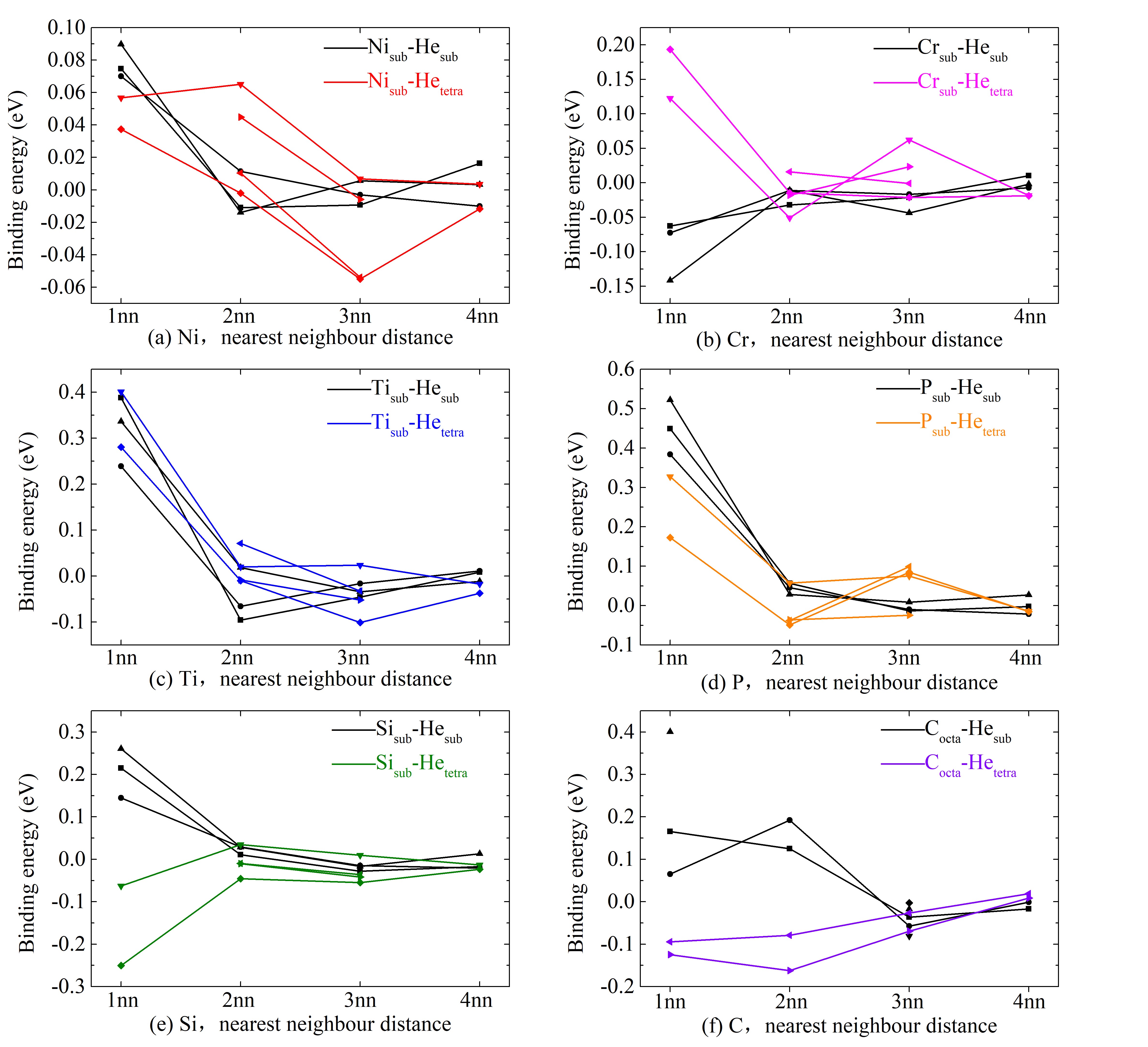}
\caption{The binding energies for solute-He pairs in afmD Fe of 1 nn to 4 nn, dark line refer to the binding energies of solute-He$_\emph{sub}$ pairs, and the color line refer to the binding energies of solute-He$_\emph{tetra}$ pairs, corresponding configurations as shown in Fig. 3 and Fig. 5.}   \label{fig5}
\end{figure*}

First, we consider He in substitutional sites. We found that the alloying element (Ti, P, Si, and C) exhibit attraction to substitutional He at 1 nn. Among these interactions between solutes and He$_\emph{sub}$, P-He$_\emph{sub}$ has the most considerable binding energy of 0.52 eV as shown in Fig. \ref{fig5}d, which can be comparable to P-He$_\emph{sub}$ of 0.54 eV in bcc Fe at 1 nn \cite{Zhang2017EffectOC}. The largest binding energies for Ti-He$_\emph{sub}$ and Si-He$_\emph{sub}$ are 0.39 eV and 0.26 eV as shown in Fig. \ref{fig5}c and e, respectively, which are both lower than that in bcc Fe \cite{Zhang2017EffectOC}. However, C-He$_\emph{sub}$ binding energy at 1 nn is 0.08 eV higher than that in bcc Fe \cite{Zhang2017EffectOC}. For the main alloying elements (Ni and Cr), Ni has a binding energy of around 0.1 eV to substitutional He showing weak attraction at 1 nn as shown in Fig. \ref{fig5}a, while Cr has repulsive interaction to substitutional He with a binding energy of -0.14 eV as shown in Fig. \ref{fig5}b, which is consistent with previous work \cite{Hepburn2013FirstPS}. We also find that the separation distance for Ti-He$_\emph{sub}$ and C-He$_\emph{sub}$ are reduced over 0.60 and 0.28 {\AA} at 1 nn under relaxation, respectively, while there is almost no displacement for other solute-He$_\emph{sub}$ pairs. Like the results at 1 nn, P, Si, and C exhibit positive binding energies to He$_\emph{sub}$ at 2 nn, in which C-He$_\emph{sub}$ has the highest binding energy of 0.19 eV. All solutes exhibit very weak repulsion to He$_\emph{sub}$ at 3 nn, and the interactions tend to disappear at 4 nn.


Then, we turn to another situation of He at interstitial sites.
Solutes Cr, Ti, and P attract interstitial He at 1 nn with the binding energy energies of 0.19 eV, 0.40 eV and 0.33 eV as shown in Fig. \ref{fig5}(b-d), respectively. While the largest binding energy for Ni-He$_\emph{tetra}$ pair is 0.07 eV at 2 nn not at 1 nn as shown in Fig. \ref{fig5}a. Especially, unlike Si-He$_\emph{sub}$ pair at 1 nn, the interaction of Si-He$_\emph{tetra}$ pair show inverse character as repulsion showing in Fig. \ref{fig5}e, which was predicted as similar interaction between Si and He in bcc Fe \cite{Zhang2017EffectOC}. The binding energy of Si-He$_\emph{tetra}$ pair is -0.25 eV at 1nn as shown in Fig. \ref{fig5}e, while there is a small positive binding energy to Si-He$_\emph{tetra}$ pair of 0.03 eV at 2 nn.
In fcc Fe, Cr repels substitutional He while it attracts interstitial He. And contrast to C-He$_\emph{sub}$ pair, there are all repulsive binding for C-He$_\emph{tetra}$ except 4 nn as shown in Fig. \ref{fig5}f. Separation distances of the configurations are increased about 0.60 {\AA} for P-He$_\emph{tetra}$, Si-He$_\emph{tetra}$, and C-He$_\emph{tetra}$ pairs under relaxation at 1nn, respectively. There are also increased separation distances for Ti-He$_\emph{tetra}$ and Ni-He$_\emph{tetra}$ pairs about 0.3 {\AA} at 1 nn under relaxation and about 0.2 {\AA} to Cr-He$_\emph{tetra}$ pair. For 3 nn and 4 nn, the variation of separation distances for all the solutes to He$_\emph{tetra}$ will be no more than 0.2 {\AA}, the binding energy of solute-He$_\emph{tetra}$ is also tend to zero at 4 nn.

We have calculated the interactions between the solute and He and got reasonable binding energies, which might help understand the contributions of solutes on precipitation capture diffused He in fcc Fe. Because of the larger binding energies between them and He, solute Ti and P do better in the point defect collector mechanism.

\section{Effects of solute on He-He pairs and their interactions}

\begin{figure*}[tbp]
\center\includegraphics[scale=0.5]{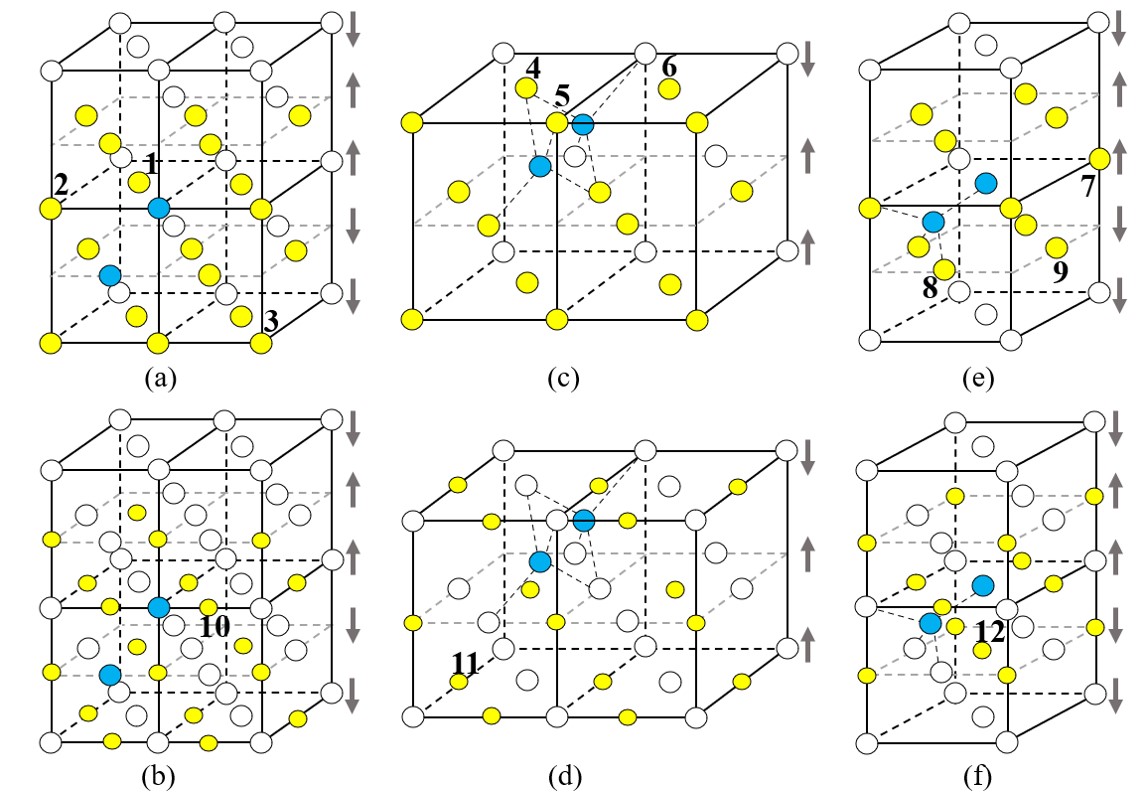}
\caption{Configurations of solute-He-He complexes, white spheres refer to the matrix Fe atoms, and blue spheres refer to He atoms. Arrows represent the local magnetic moments. For the addition of solute, yellow spheres refer to the studied positions, and the spheres figured out by numbers refer to the lowest energy configuration. In detail, for substitutional solutes (Ni, Cr, Ti, P, Si), (a) solute-He$_\emph{sub}$-He$_\emph{sub}$ complexes, the position of 1 is occupied by Ti or P or Si, the position of 2 is occupied by Ni and the position of 3 is occupied by Cr. (c) Solute-He$_\emph{tetra}$-He$_\emph{tetra}$ complexes, the positions of 4 is occupied by Ti or P, Cr occupies the position of 5, and Ni or Si occupies the position of 6. (e) Solute-He$_\emph{sub}$-He$_\emph{tetra}$ complexes, the position of 7 is occupied by Cr, Ti occupies the position of 8, Ni or P or Si occupies the position of 9. For octahedral solute C, (b) C-He$_\emph{sub}$-He$_\emph{sub}$ complexes, (d) C-He$_\emph{tetra}$-He$_\emph{tetra}$ complexes. (f) C-He$_\emph{sub}$-He$_\emph{tetra}$ complexes.} \label{fig6}
\end{figure*}

We have obtained solute interactions on a single He from the above prediction.
To know the growth of He bubbles in the presence of solute, we then consider the self-trapping with the addition of solute at an atomic scale. According to the calculated interaction between He-He pairs,
we have known the relatively favourable He-He pairs on energy. The stable He-He pairs used in the following calculations are He$_\emph{sub}$-He$_\emph{sub-1c}$, He$_\emph{sub}$-He$_\emph{tetra-1a}$ and He$_\emph{tetra-2d}$-He$_\emph{tetra-3d}$ as shown in Fig. \ref{fig6} labeled as blue atoms. Then, we built possible neighbourhood positions for solute to He-He pairs to find the possible stable solute-He-He complexes.
Based on these complexes, we have calculated the effects of solute on the combination of two He atoms, as well as the interactions between solute and He-He pairs.

\begin{figure*}[tbp]
\center\includegraphics[scale=0.34]{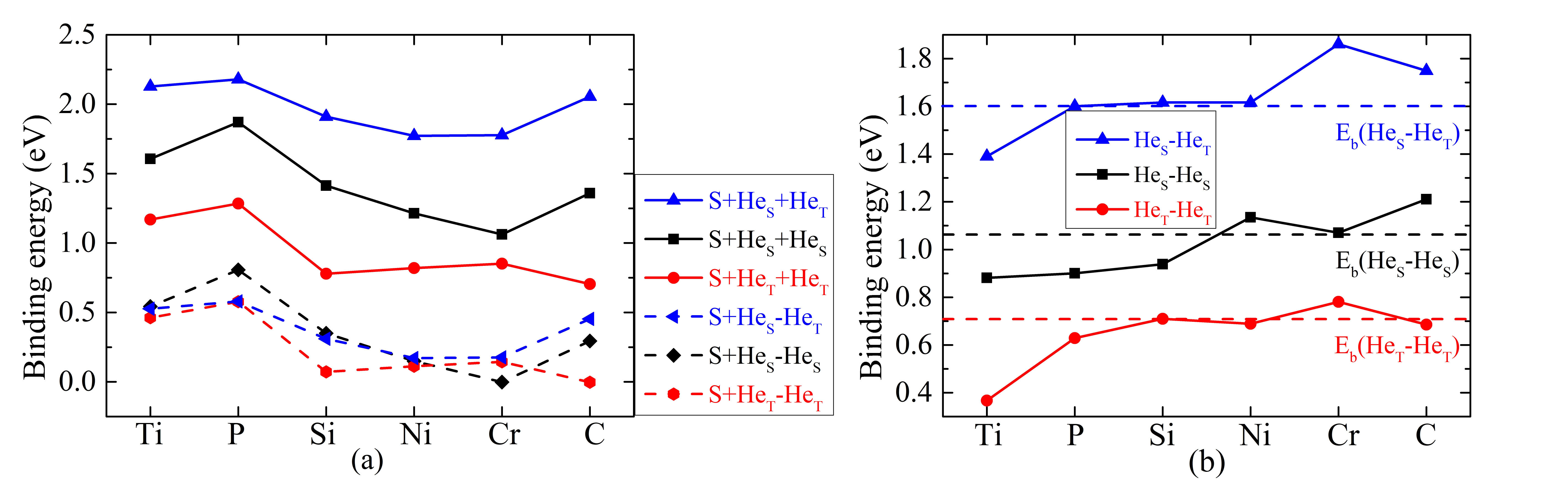}
\caption{(a) The total binding energies of solute-He-He complexes and the binding energies between solute and He-He pair. (b) The solid line refers to the binding energies of He-He pairs when solute exists and the dashed line refers to the binding energies of He-He pairs in pure Fe.} \label{fig7}
\end{figure*}
First of all, we show the interactions of solute and He-He pairs.
Before calculating the interactions, we determine the stable of the solute-He-He clusters. As shown in Fig. \ref{fig7}a (solid line), the positive total binding energies for solute-He-He clusters show the stability of these clusters. We have calculated the binding energies between all the solutes and three mentioned He-He pairs.
For solutes Ti, P, Si, and C, the largest binding energy between every solute and the three He-He pairs is 0.54 eV, 0.81 eV, 0.35 eV and 0.45 eV, respectively, respectively, as shown in Fig. \ref{fig7}a (dashed line).
However, for main elements Ni and Cr, the largest binding energies between solute and He-He pairs are 0.17 eV and 0.18 eV, respectively, which are much lower than that of Ti, P, Si, and C. Similar to the interactions of solute and He, the binding energy of trace alloying elements to He-He pair is higher than that of major alloying elements to He-He pair.

\begin{figure}[h]
\center\includegraphics[scale=0.5]{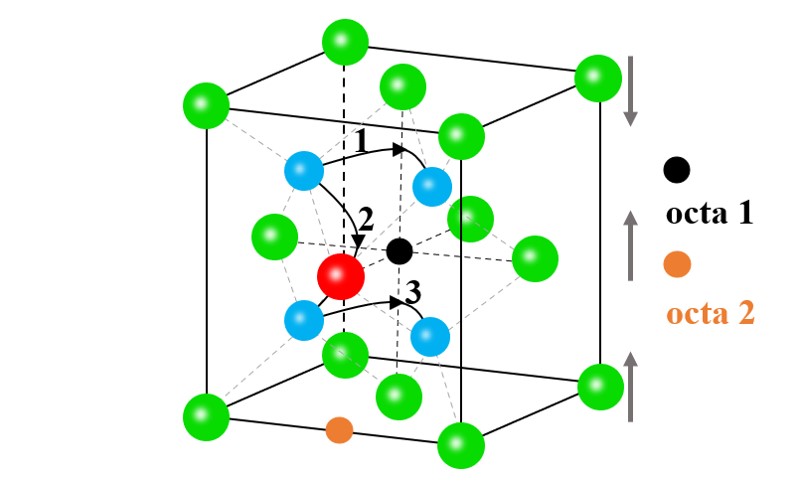}
\caption{Possible migration paths for interstitial He in afmD Fe lattice. Paths are shown for 1 nn jumps from initial to final tetra positions (blue spheres). Green spheres are referred to Fe atoms; red sphere refers to Fe or substitutional solute atom, black circle refers to octahedral site 1 (octa1), orange circle refers to octahedral site 2, which is one of the nearest neighbour octa sites to octa 1, arrows indicate the local moments.} \label{fig8}
\end{figure}

Then, we consider the stability of He-He pairs affected by solutes to elucidate the accumulation of He in fcc iron with solute addition at the early state. As shown in Fig. \ref{fig7}b,
Ti is visibly harmful to He self-trapping. The binding energy of He$_\emph{sub}$-He$_\emph{sub}$ and He$_\emph{sub}$-He$_\emph{tetra}$ pairs are reduced by 0.18 eV and 0.12 eV with Ti addition, in particular, the binding energy of He$_\emph{tetra}$-He$_\emph{tetra}$ pair is reduced by nearly half relative to 0.71 eV. The addition of P will also reduce the binding energies of He$_\emph{sub}$-He$_\emph{sub}$ and He$_\emph{tetra}$-He$_\emph{tetra}$ pairs by 0.16 eV and 0.08 eV, but do not affect He$_\emph{sub}$-He$_\emph{tetra}$ pair. For Si, it can only affect the stability of He$_\emph{sub}$-He$_\emph{sub}$ pair, reducing the pair binding energy by 0.13 eV. However, C is beneficial for He self-trapping, the binding energies of He$_{sub}$-He$_{sub}$ and He$_{sub}$-He$_{tetra}$ pairs are both increased by 0.15 eV but have almost no effect on He$_\emph{tetra}$-He$_\emph{tetra}$ pair. For main alloying elements, Ni increased the binding energy of He$_\emph{sub}$-He$_\emph{sub}$ slightly by 0.07 eV and does not influence He$_\emph{tetra}$-He$_\emph{tetra}$ and He$_\emph{sub}$-He$_\emph{tetra}$ pairs. However, the addition of Cr increases the binding energies of He$_{tetra}$-He$_{tetra}$ and He$_{sub}$-He$_{tetra}$ pairs by 0.26 eV and 0.07 eV.
Summarized above, Ti, P, and Si inhibit self-trapping to varying degrees. At the same time Cr and C promote self-trapping, and Ni has little effect on self-trapping.

\section{Effects of solutes on interstitial He diffusion}
The migration of He is an essential process for accumulating He atoms which can form He bubbles. To date, many efforts have been devoted to studying He migration behaviour in V, bcc-Fe, Al, Pd, and W alloys \cite{yang2008ab,zeng2009first,fu2005ab,Li2016EffectsOC,Becquart2006MigrationEO,Zhang2011StabilityAM}.
Here we want to calculate the migration energy barrier of interstitial He and the effects Ni, Cr, Ti, P and Si on the energy barrier of He migration in pure fcc Fe. We have known that the most stable interstitial He is in the tetrahedral site. Therefore, we considered a single interstitial He migrates between adjacent tetrahedral sites, which can diffuse along three paths, as shown in Fig. \ref{fig8}. The path is not a simple straight line but a curve. To determine possible paths, we have relaxed a supercell with He initially located in the octahedral site and found that the final stable He site is not the octahedral site. Still, it displaces a little to occupy an off-centre octahedral site, a promising intermediate state for interstitial He migration. We have performed CINEB calculations for He migration along these paths to find the energy barrier for corresponding transition states.

\begin{figure}[h]
\center\includegraphics[scale=0.3]{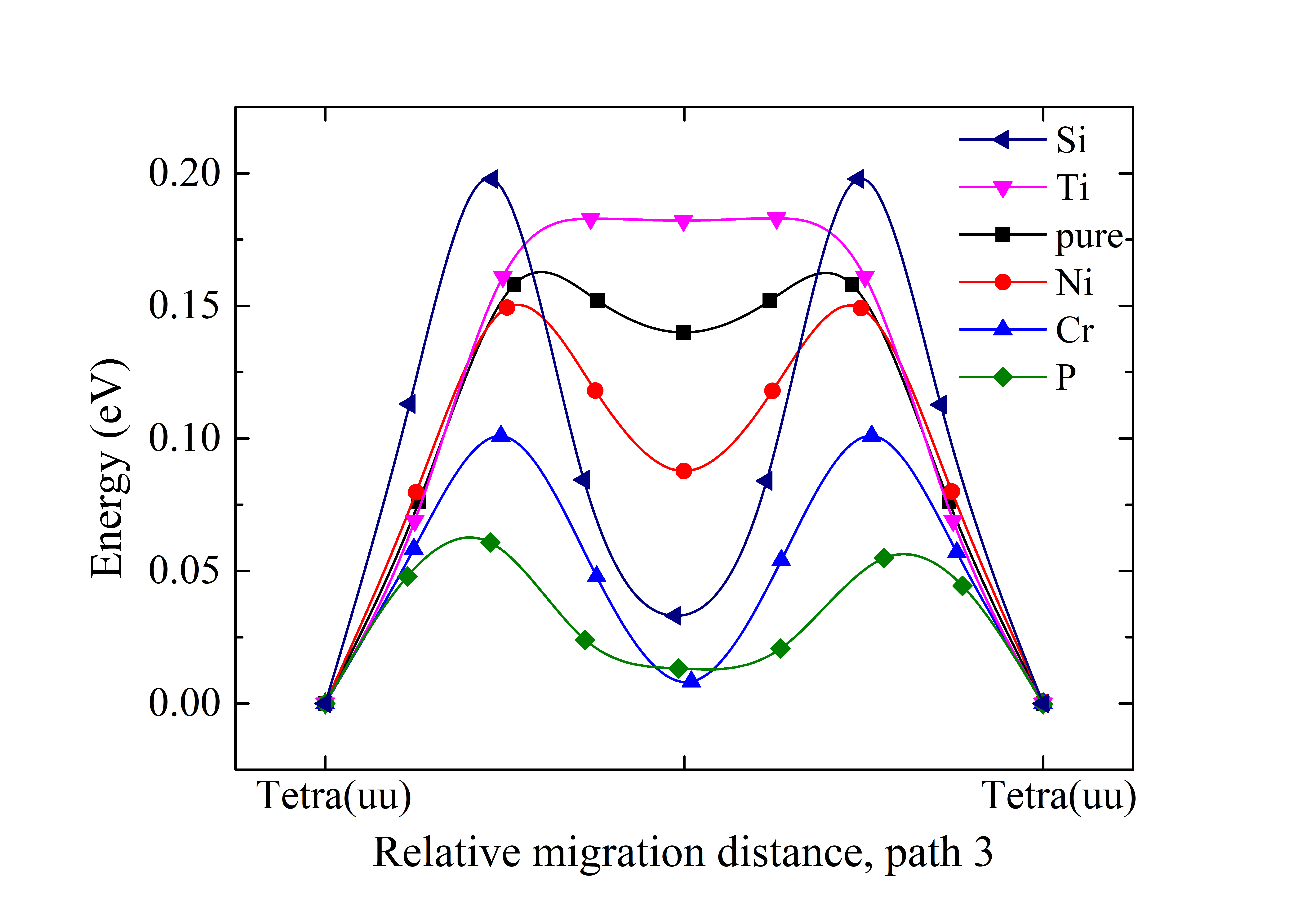}
\caption{Migration energy barriers of interstitial He along path 3 with and without solutes.} \label{fig9}
\end{figure}

Firstly, we have determined the path that is the possible one among the three paths. We find the migration energy barrier of an interstitial He in pure fcc Fe is 0.16 eV along the path 3, which is the lowest energy barrier among three paths as shown in Fig. \ref{fig9} and Fig. S3, and the result is consistent with previous work of 0.16 eV \cite{Hepburn2013FirstPS}. Such an interstitial He migration barrier is also reasonable in other metals, such as 0.04 eV in Pt \cite{Cao2016MigrationOH}, 0.06 eV in bcc Fe \cite{Li2016EffectsOC}, 0.07 eV in W \cite{Becquart2006MigrationEO} and V \cite{Zhang2011StabilityAM}, 0.08 eV in Cu \cite{Cao2016MigrationOH}, 0.10 eV in fcc Al \cite{yang2008ab}, 0.13 eV in Ni \cite{Hepburn2013FirstPS}, and 0.15 eV in Pd \cite{Cao2016MigrationOH}. We apply path 3 to investigate the effect of solutes on the migration of interstitial He in the following.

Then we calculated the migration energy barriers of He along path 3 with a single arbitrary solute added. The results show that interstitial He migration energy barrier is 0.15 eV with Ni addition, which is comparable to that of pure fcc Fe.
While Cr decreases the migration energy barrier to 0.10 eV, it is consistent with the behaviour of Cr, which reduces the migration energy barrier of interstitial He by half in bcc Fe \cite{Li2016EffectsOC}. Elements Ti and Si increase interstitial He migration energy barrier to 0.18 eV and 0.20 eV, are higher than 0.16 eV in pure fcc Fe. However, P considerably decreases the energy barrier to 0.06 eV.

We found that migration intermediate states of path 3 are kept in the (100) plane, but we show that different solute results in another direction of diffusion. For Ni, Ti, and Si addition, the direction of migration is toward octa 1, while for Cr and P addition, migration is toward octa 2, as shown in Fig. \ref{fig8}.
However, as mentioned before, the site of intermediate migration state is not exactly at the octa 1 or octa 2, but new sites with little displacing from the centre of octahedral sites. This unique property belongs to He rather than other solutes like H, C, and N.

In short, the migration energies barrier of interstitial He can be raised by Ti and Si addition, while the addition of Cr and P are beneficial for He diffusion, Ni almost does not affect the migration of interstitial He.
The migration energy of interstitial He in the presence of impurities can be a starting point to estimate effective diffusion coefficients in mean-field approaches providing frameworks to evaluate materials swelling.

\section{Conclusion}
First-principles calculations have been performed to understand the effects of alloying elements (Ni, Cr, Ti, P, Si, C) on the behaviours of He in fcc iron. We found strong attractions between two He atoms, and the binding energies are 0.71 eV, 1.06 eV and 1.60 eV for He$_{tetra}$-He$_{tetra}$, He$_{sub}$-He$_{sub}$ and He$_{sub}$-He$_{tetra}$ pairs, respectively. All solutes can attract substitutional He except Cr. While the interactions with interstitial He, solutes Ti, P, Ni, and Cr can also attract, but Si and C show inverse properties of interacting. For the interactions of He-He pairs, Ti, P, and Si weaken He self-trapping, Cr and C are beneficial for He self-trapping, and Ni affects the stability of He-He pairs slightly. The most stable interstitial He is in the tetrahedral site and migrate with the energy barrier of 0.16 eV. The energy barriers of migration of interstitial He are increased to 0.18 eV, 0.20 eV with Ti and Si, but decreased to 0.10 eV and 0.06 eV with addition of Cr and P, respectively.
Hence, we could conclude that Ti decreases, but Cr increases the diffusion and self-trapping of He in fcc iron. Further challengeable work is needed to solve synergy solutes on He behaviours in fcc iron or even austenitic steel.

\section*{acknowledgments}
The authors are grateful for financial support from National Natural Science Foundation of China (General program 51671086 and U1867217) and Hunan Provincial Education Department (Key project 19A324).

\section*{References}
\bibliography{He-He}

\end{document}